\documentclass[12pt]{article}
\usepackage{a4}
\usepackage{cite}
\usepackage{epsfig}
\usepackage{amsmath}
\usepackage{amssymb}

\newcommand{\de}{\partial}
\newcommand{\lsi}{\raise0.3ex\hbox{$<$\kern-0.75em\raise-1.1ex\hbox{$\sim$}}}
\newcommand{\gsi}{\raise0.3ex\hbox{$>$\kern-0.75em\raise-1.1ex\hbox{$\sim$}}}

\newcommand{\gsim}{\mathop{\gsi}}
\newcommand\gd\delta
\newcommand{\tb}{\tan\!\beta}

\begin{document}
\begin{flushright}
IEM-FT-206/00\\
IFT-UAM/CSIC-00-33\\
HD-THEP-00/59\\
hep-ph/0012077\\
\end{flushright}
\begin{centering}

{\bf\Large Bubble Wall Velocity in the MSSM\footnote{To appear in the Proceedings of SEWM2000, Marseille, June 14-17, 2000 }}
\vspace{1cm}

P. John\\

{\em Instituto de Estructura de la Materia, CSIC, Serrano 123, 28006
Madrid, Spain\\
E-mail:john@makoki.iem.csic.es}

\vspace{1cm}

M.G. Schmidt\\

{\em Institut f\"ur Theoretische Physik, Philosophenweg
16, D-69120 Heidelberg, Germany}\\	
{\em E-mail:M.G.Schmidt@thphys.uni-heidelberg.de }

\end{centering}
\vspace{1cm}
\abstract{We compute the wall velocity in the MSSM with
$W$, tops and stops contributing to the friction.  In a wide range of
parameters including those which fulfil the requirements of
baryogenesis we find a wall velocity of order $v_w\approx 10^{-2}$
much below the SM value.  }

\section{Bubble Wall Equation of Motion in the MSSM}
Energy conservation leads to the equations of motion of an electroweak
bubble wall interacting with a hot plasma of particles:
\begin{eqnarray}
\square h_1 + \frac{\de V_T(h_1,h_2)}{\de h_1} + \sum_i\frac{\de
m_i^2}{\de h_1}\int\frac{d^3p}{(2\pi)^32E_i}\delta f_i(p,x) &=& 0,\\
\square h_2 + \frac{\de V_T(h_1,h_2)}{\de h_2} + \sum_j\frac{\de
m_j^2}{\de h_2}\int\frac{d^3p}{(2\pi)^32E_j}\delta f_j(p,x) &=& 0.
\end{eqnarray}
where $f_i=f_{0,i}+\delta f_i$ is the distribution function for a
particle species in the heat bath\cite{MP12,John6}. We have to sum over all particle
species $i$.  The distribution function is divided up into equilibrium
part $f_{0,i}$ and out-of-equilibrium part $\delta f_i$.
The equilibrium part has been absorbed into the equilibrium
temperature dependent effective potential $V_T(h_1,h_2)$. 

In the following we will restrict ourselves to late times leading to a
stationarily moving domain wall where the friction stops the bubble
wall acceleration. This is a reasonable assumption for the late stage
of the bubble expansion where baryogenesis takes place. 
\section{Fluid Equations}
\label{sec:fleqs}
The deviations $\delta f_i$ from the equilibrium population density
are originating by a moving wall. They are derived by Boltzmann
equations in the fluid frame:
\begin{equation}
d_tf_i\equiv \de_tf_i + \dot x \frac{\de}{\de x} f_i + \dot p_x \frac{\de}{\de_{p_x}} f_i = -C[f_i],\label{B1}
\end{equation}
with the population density $f_i$ and energy $E=\sqrt{p_x^2 +m^2(x)}$.
$C[f_i]$ represents the scattering integral. The classical (WKB)
approximation is valid for $ p\gg {1}/{L_w} \quad\mbox{(``thick
wall'')}.$ For particles with $E,p \gsim gT$ this should be fulfilled.
Infrared particles are supposed not to contribute to the friction.
This is a crude approximation and there are additional
contributions\cite{Moorewall2000} which further lower the wall
velocity. In the MSSM the wall thickness $L_w$ is of order
$15/T$--$40/T$, as found in~\cite{SecoNum,John3}, and $L_w\gg 1/T$ is
fulfilled. Those particles which couple very weakly to the Higgs are
denoted as ``light particles''.  Particles coupling strongly to the
Higgs are heavy in the Higgs phase and therefore called
``heavy''. ``superheavy'' particles as the ``left handed'' stops do
not appear in our calculation besides their remnants in the effective
potential.  We treat as ``heavy'' particles top quarks, (right handed)
stops, and W bosons. The Higgses are left out. Further contributions
produce an even smaller velocity.  We assume now that the interaction
between wall and particle plasma is the origin of small perturbations
from equilibrium. We will treat perturbations in the temperature
$\delta T$, velocity $\gd v$ and chemical potential $\delta\mu$ and
linearize the resulting fluid equations. Then the full population
density $f_i$ of a particle species $i$ in the fluid frame is given by
\begin{equation}
f_i= \frac{1}{\exp\left\{\frac{(E+\gd_i)}{T} \right\}\pm 1}
\end{equation}
where we have generally space dependent perturbations $\gd_i$ from
equilibrium. In principle one must include perturbations  for
each particle species. A simplification is to treat all the ``light''
particle species as one common background fluid.  This background
fluid obtains common perturbations $\gd v_{bg}$ in the velocity and
$\gd T_{bg}$ in the temperature. This leads us to
\begin{equation}
\gd_i= -\left[ \delta\mu_i + \frac{E}{T}(\gd T_i + \gd T_{bg}) + p_x(\delta v_i+\delta v_{bg})\right]
\label{gd}
\end{equation}
for the ``heavy'' particles. The spatial profiles of all these
perturbations depend on the microscopic physics.  We treat particles
and antiparticles as one species neglecting CP violation which is a
minor effect on the friction. It were, of course, important for the
calculation of the baryon asymmetry.

Since the perturbations are Lagrangian multipliers for particle
number, energy, and momentum, we can expand (\ref{B1}) to a set of three
equations, called ``fluid equations'', coupled by the collision term
$C[\gd \mu, \gd T,gd v]$.  Performing the integrals\cite{John6,MP12} leads to
the general pattern
\begin{eqnarray}
\int\!\!\!\frac{d^3p}{(2\pi)^3T^2} C[f]\!=\!\delta\mu\Gamma_{\mu_1}\!\!\!+\!\gd T \Gamma_{T_1},&\ &\int\!\!\!\frac{d^3p}{(2\pi)^3T^3}E C[f]\!=\!\delta\mu\Gamma_{\mu_2}\!\!+\!\gd T \Gamma_{T_2},\nonumber\\
\int\frac{d^3p}{(2\pi)^3T^3}p_x C[f] &=& \gd vT\Gamma_v,\label{gammadef}
\end{eqnarray}
For a stationary wall we can use $\de_t\delta_i\to
\gamma_wv_w\delta'_i,$ and $ \de_z\delta_i\to \gamma_w\delta'_i$,
where the prime denotes the derivative with respect to $z=\gamma_w(x-v_wt)$. Our  equations are  similar to those in \cite{MP12} but there are
important additional terms. 

For each heavy particle species in the plasma we have three fluid
equations.  The final form of the fluid
equations can be written in a matrix notation:
\begin{equation}
\mathbb{A}\delta'+\Gamma\delta=F,
\label{fluidfinal}
\end{equation}
where $\Gamma=\Gamma_0+1/\bar{c}_4\mathbb{M}.$
The matrices $\mathbb{A}$, $\Gamma$, $\Gamma_0$, and $\mathbb{M}$ can
be found in \cite{John6}.  The number $\bar{c}_4$ is the heat capacity
of the plasma $\bar{c}_4=78 c_{4-}+37c_{4+}$ including light quarks,
leptons, and sleptons in the plasma.  The perturbations are combined in a
vector $\delta$, the driving terms are combined in the vector $F$. The
driving term containing $(m^2)'$ can be split up into different
contributions
\begin{equation}
(m^2)' = \frac{\de m^2}{\de h_1}h_1' + \frac{\de m^2}{\de h_2}h_2'.
\label{msquared}
\end{equation} 
The vectors $\delta$ and $F$  for $k$ particle
species read (index $x$ denotes $+$ or $-$, for fermions and bosons,
respectively, for the $i$th particle):
\begin{eqnarray}
\gd=\begin{bmatrix}
\gd\mu_1 & \gd T_1 & T\gd v_1 & \ldots &\gd \mu_k&\gd T_k & T\gd v_k\end{bmatrix}^T
,\\
 F=\frac{\gamma_wv_w}{2T}\begin{bmatrix}
c_{1\pm}(m_1^2)'&c_{2\pm}(m_1^2)'& 0 &\ldots &&c_{1\pm}(m_k^2)'&c_{2\pm}(m_k^2)'&0\end{bmatrix}^T.
\end{eqnarray}
where  $c_{1\pm}$, $c_{2\pm}$  denote the  fermionic(+) or bosonic(-)
statistical factors, respectively, defined through
$
c_{n\pm}= \int {E^{n-2}}/{T^{n+1}}(f_0(\pm)^\prime){d^3p}/{(2\pi)^3}.\label{definitionc}
$
Eq.~(\ref{fluidfinal}) has to be solved for $\delta$. To a first approximation, neglecting $\delta'$, we obtain 
$\delta=\Gamma^{-1}F$. Then, including (right-handed) stop-, top- and $W$ particles, the equations of motion can be rewritten  in the fluid picture as
\begin{eqnarray}
h_1^{\prime\prime}-V'_T=\eta_1\gamma_wv_w\frac{h_1^2}{T}h_1',
\quad h_2^{\prime\prime}-V'_T=\eta_2\gamma_wv_w\frac{h_2^2}{T}h_2'.\label{weqmo}
\end{eqnarray}
with slightly\! $\tb$-dependent friction constants $\eta_{1,2}\!\!=\!\!T/4
G_{1,2}\Gamma^{-1}\tilde{F}_{1,2}\label{eta1}$, with constant
vectors\cite{John6}\! $\tilde{F}_{1,2}$ and $G_{1,2}$.  Perhaps $\gd'$
is not negligible, so we have to solve~(\ref{fluidfinal})
numerically. We compare both resulting velocities later on in
Fig.~\ref{fig:velo}.
\section{Wall Velocity in the MSSM}
\label{sec:wv}
\begin{figure}
\hspace{1.5cm}
{\epsfxsize=13pc{\epsffile{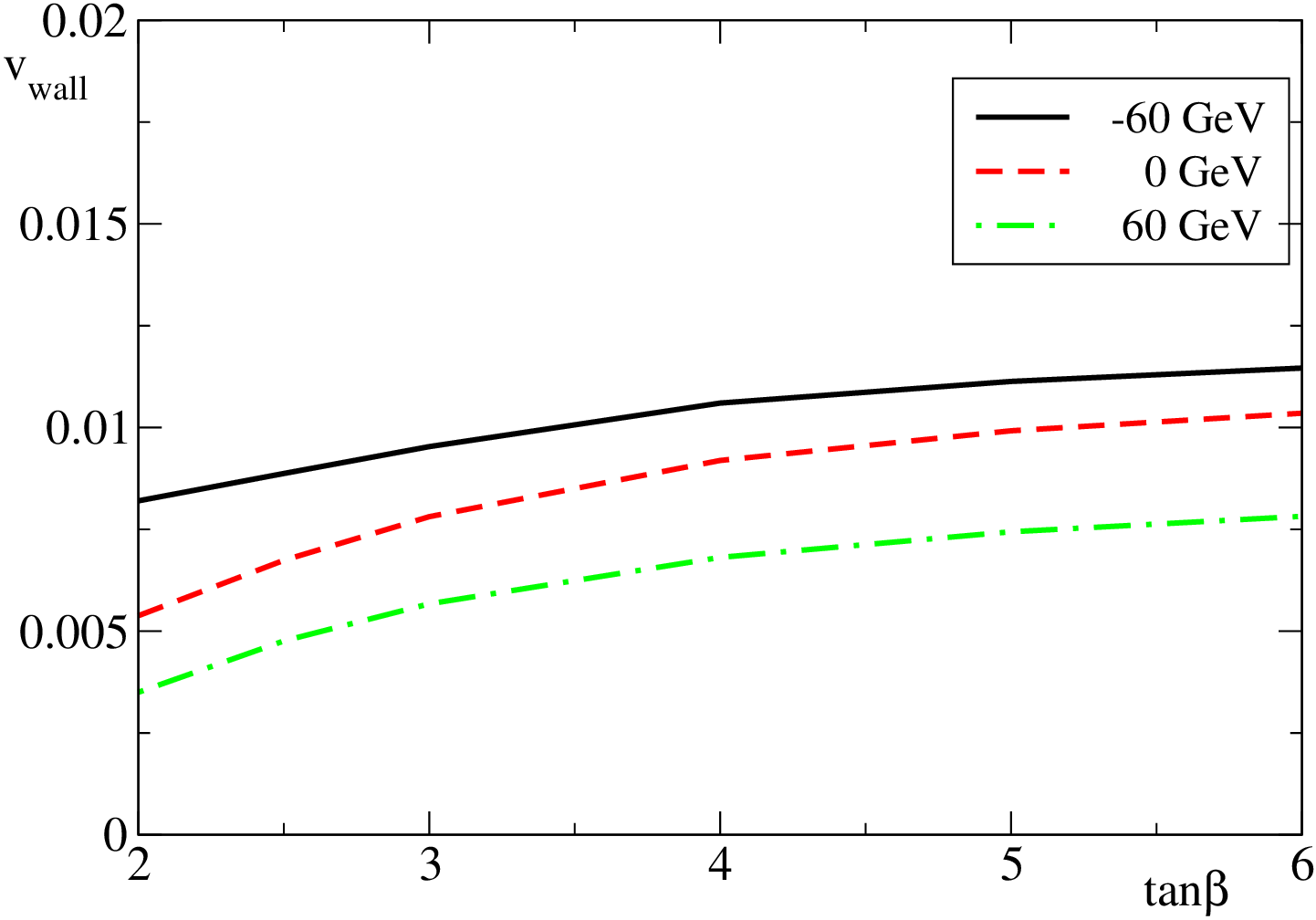}}}
\hspace{1.2cm}
{\epsfxsize=13pc{\epsffile{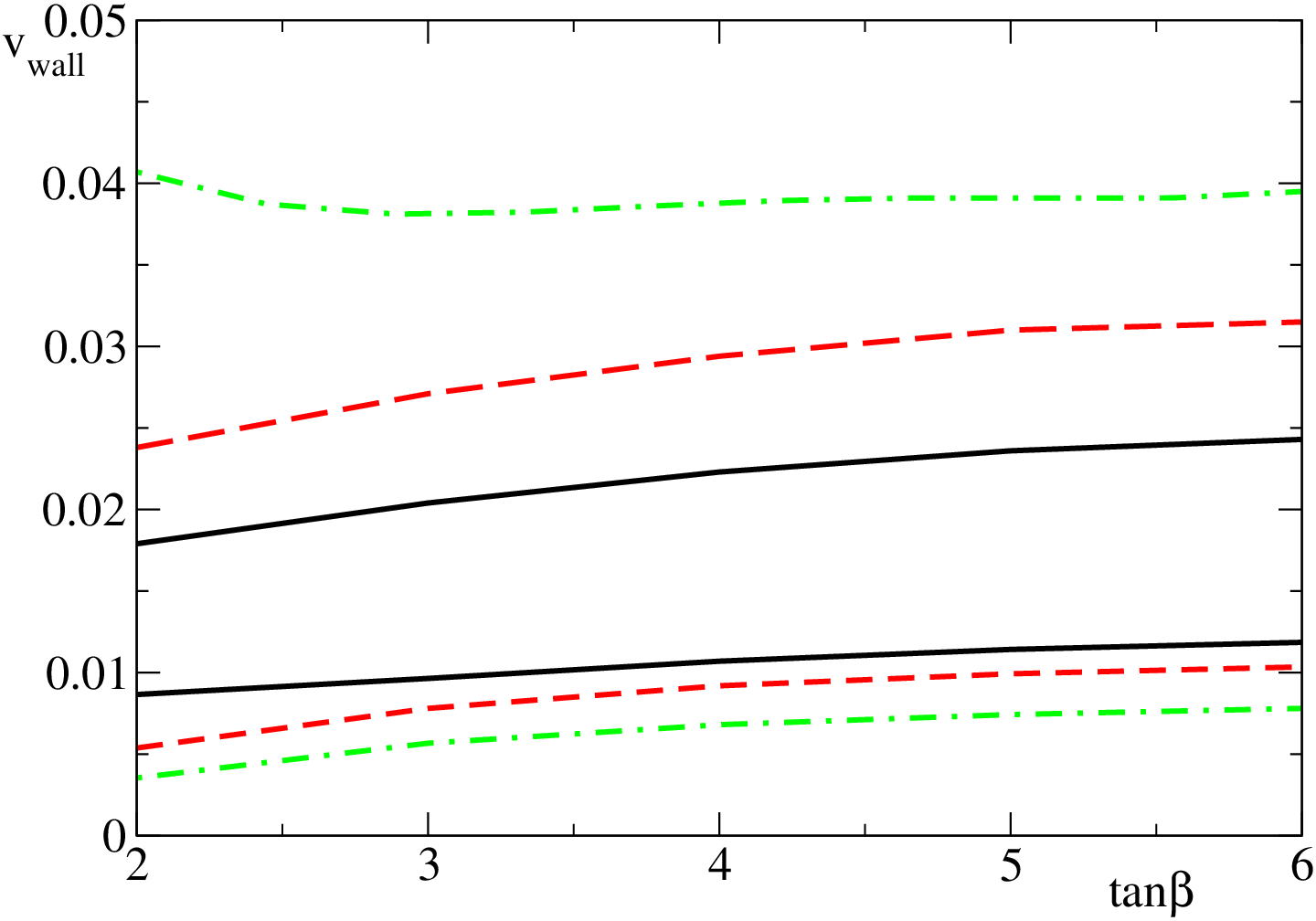}}}
\caption{Wall velocity in dependence on the parameter $\tb$ for
$m_U=-60,0,60$~GeV; left: $\delta'=0$, right: the same plus velocities
for $\delta'\neq 0$ resulting from the full solution of
(\ref{fluidfinal}).}
\label{fig:velo}
\end{figure}

In order to solve eqs.~(\ref{weqmo}) we derive a virial theorem, based
on the necessity that for a stationary wall the pressure to the wall
surface is balanced by the friction. The pressure on a free bubble
wall can be obtained from l.h.s. of the equations of motion
(\ref{weqmo}) integrated with $h_{1,2}'$, e.g. 
\begin{equation}
p_1=\int_0^{\infty}\left(h_1^{\prime\prime}-\frac{\de V_T}{\de h_1}\right)h_1'dx=\Delta V_T=\int\eta_1\gamma_wv_w\frac{h_1^2}{T}(h'_1)^2.
\end{equation}
$\Delta V_T$ is the difference in the effective potential values at
the transition temperature $T_n$, which is basically the nucleation
temperature. Both of the Higgs fields develop friction terms and we
have to add the pressure contributions. In the MSSM the
approximation\cite{SecoNum,John3} to the solution by a kink $h(x)={h_{crit}}/{2}\left(1+\tanh({x}/{L_w})\right)$ is a rather good choice. With $h_2=\sin\!\beta$ and $h_1=\cos\!\beta$ we are left with 
\begin{equation}
\gamma_wv_w=\frac{20L_w}{h_{crit}^4}\frac{\Delta V_T(T_n) T_n}{\sin^4\!\beta(\eta_2+\eta_1\cot^4\!\beta)}.
\label{master}
\end{equation}
The missing numbers for $L_w$, $h_{crit}$, $T_n$, and $\Delta V(T_n)$
can be independently determined with methods described in
\cite{SecoNum,John3}. We used the 1-loop resummed effective potential.
The diagram is calculated for $m_Q=2TeV$, and $m_A=400GeV$. We find a
strong phase transition with $v/T=.95$ at 1-loop level for
$m_U=-60GeV$ at $\tb=2.0$ and $A_t=\mu=0$ (no stop mixing). At 2-loop
level it is even much larger\cite{litestop} permitting larger $\tb$
for the same strength. Also with mixing in the stop matrix this allows
to comply the experimental lower bound on the Higgs mass. In
Figure~\ref{fig:velo} we show $v_W$ vs. $\tb$ for three values of
$m_U=-60,0,+60$GeV at zero mixing.  Very heavy stops should decouple
more and more which would lead to increasing wall velocities
again. This behaviour is reproduced with the full numerical solution
to (\ref{fluidfinal}), see Fig.~\ref{fig:velo}.  Nevertheless for
increasing $m_U^2$ the used approximations become worse and
corrections of order ${\cal{O}}(m^2/T^2)$ are important. Our
calculations are done with massless outer legs only taking into
account changing plasma masses.

\begin{figure}
\hspace{1.5cm}
{\epsfxsize=13pc{\epsffile{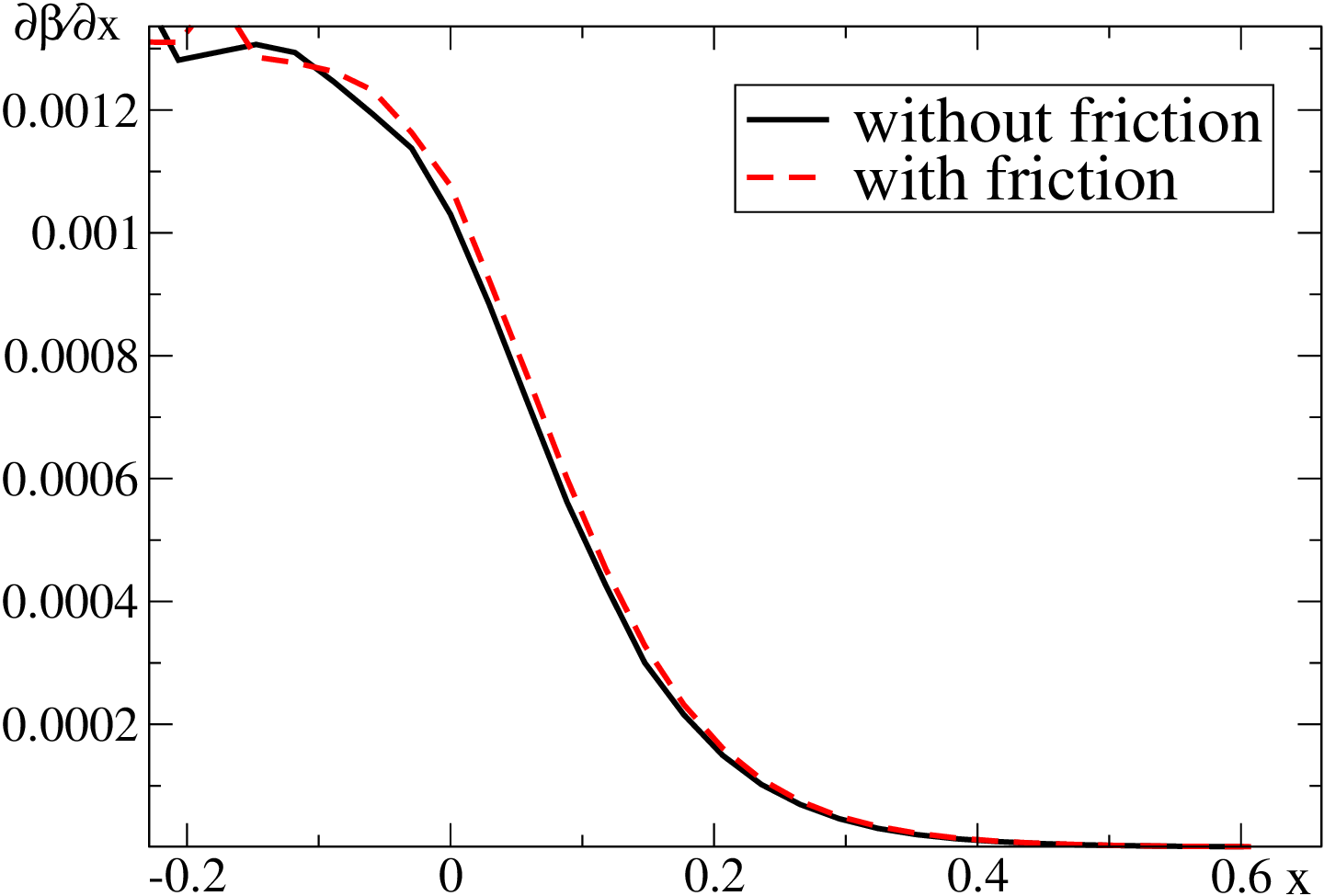}}}
\hspace{1.2cm}
{\epsfxsize=13pc{\epsffile{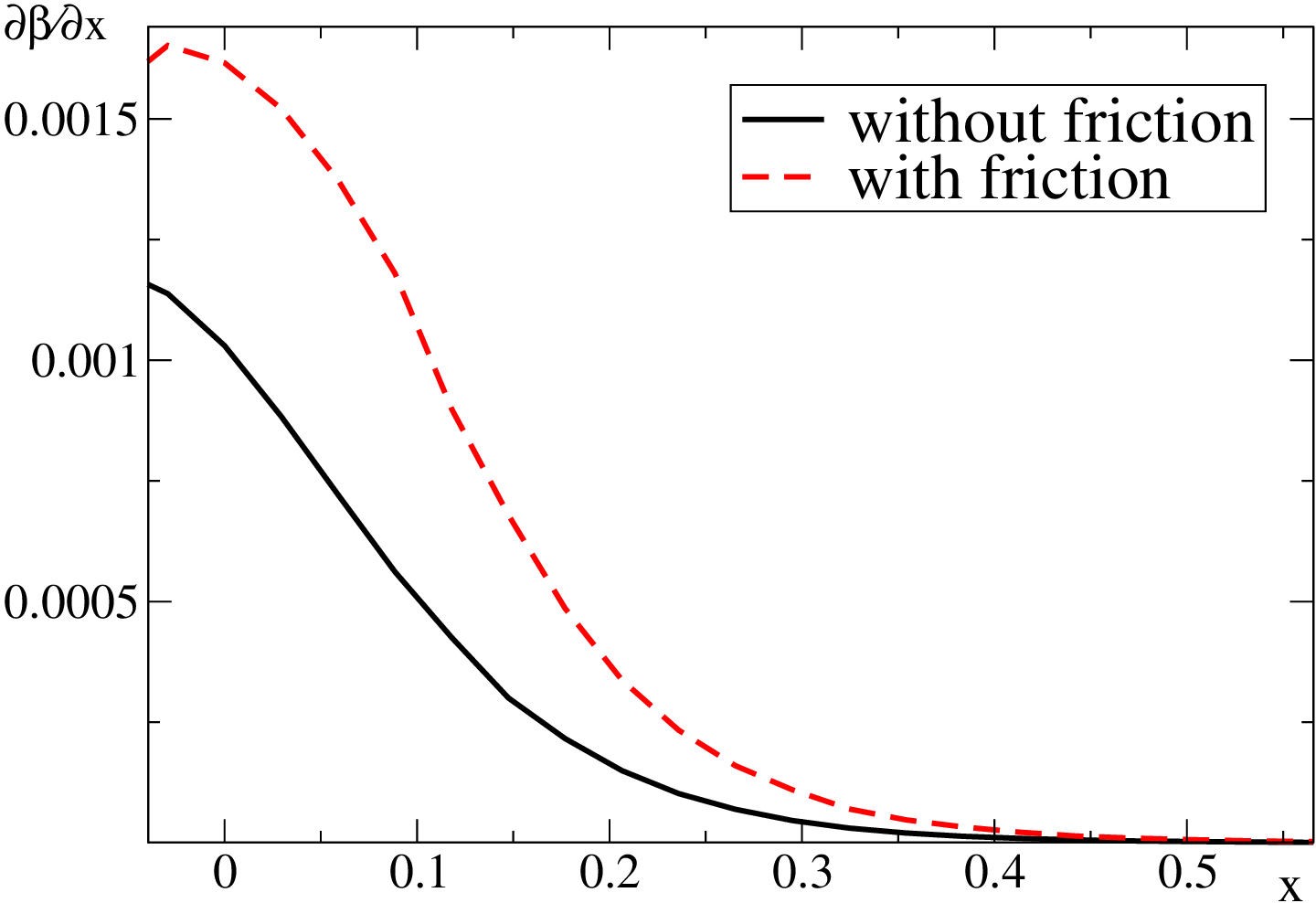}}}
\caption{$\beta(x)$ for $m_U^2=-(60GeV)^2$, $m_Q=2$TeV, $m_A=400$GeV, $\tb=2$, and no mixing. The back reaction of the friction is negligible small (left). Only artificially setting $\eta_2$ two orders of magnitude larger leads to sizable effects (right).}
\label{fig:wallbeta}
\end{figure}

Due to the effective potential which couples the equations of motion
there may be back-reaction of the different friction contributions to
$\de\beta/\de z$ leading to a change in
$\Delta\beta=\mbox{max}(\de\beta/\de z)$.  A larger $\Delta\beta$ were
highly welcome to obtain a larger baryon asymmetry. This becomes even
more important since we realized\cite{John4} that in the MSSM
transitional CP violation does not occur. Therefore we must exploit
the explicit phases which may nevertheless be strongly restricted by
experimental bounds. The determination of $\Delta\beta$ can be done
numerically by solving eqs.~(\ref{weqmo}) with extensions of known
methods\cite{John3}. But only for artificially large friction we obtain
sizable effects (see Fig.~\ref{fig:wallbeta}).

\noindent{\bf Acknowledgements} We thank G.~Moore for useful
discussions. This work was partly supported by the TMR network {\em
Finite Temperature Phase Transitions in Particle Physics}, EU contract
no.\ FMRX-CT97-0122.

\begin{thebibliography}{99}

\bibitem{MP12}
G.~Moore, T.~Prokopec, {\em Phys. Rev.} {\bf D52} (1995)  7182.

\bibitem{John6} P.~John,\! M.G.~Schmidt, hep-ph/0002050,\! revised\! version\! and\! refs.\! therein.

\bibitem{Moorewall2000} G.~Moore, JHEP 0003:006,2000.

\bibitem{SecoNum}
J.~Moreno, M.~Quir\'os, M.~Seco, {\em Nucl. Phys.} {\bf B526} (1998) 489;
J.~Cline, G.~Moore, {\em Phys. Rev. Lett.} {\bf  81} (1998) 3315;
J.~Cline, G.~Moore, G.~Servant, {\em Phys. Rev.}{\bf D60} (1999) 105035.

\bibitem{John3} P.~John, {\em Phys. Lett.} {\bf B452} (1999) 221; P.~John, Proc. of {\em SEWM 98}. 

\bibitem{litestop}
D.~B{\"o}deker, P.~John, M.~Laine, and M.G.~Schmidt, {\em Nucl.  Phys.} {\bf B497} (1997) 387; J.~Espinosa, M.~Quir\'os, and F.~Zwirner, {\em Phys. Lett.}  {\bf B307} (1993) 106; 
J.~Espinosa, {\em Nucl. Phys.} {\bf B475} (1996)  273; 
M.~Carena, M.~Quir\'os, and C.E.M~Wagner, {\em Nucl. Phys.} {\bf  B524} (1998) 3; 
M.~Losada, {{\tt hep-ph/9905441}}; 
M.~Laine and K.~Rummukainen, {\em Nucl. Phys.} {\bf B535} (1998) 423; 
ibid., {\em Phys. Rev. Lett.} {\bf 80} (1998) 5259.

\bibitem{John4}
S.~Huber, P.~John, M.~Laine, M.~Schmidt, {\em Phys. Lett.} {\bf B475} (2000) 104.

\end{thebibliography}
\end{document}